\documentclass[aps,pre,reprint,footinbib,showkeys,superscriptaddress]{revtex4-2}
\usepackage[utf8x]{inputenc}
\usepackage{ucs}
\usepackage{graphicx}
\usepackage{color}
\usepackage{amsmath}
\usepackage{lipsum}

\usepackage[inline]{trackchanges}
\addeditor{AN}
\addeditor{HD}

\begin{document}
\title{Quantifying active and resistive stresses in adherent cells.}

\author{H\'el\`ene Delano\"e-Ayari}
\email{helene.delanoe-ayari@univ-lyon1.fr}
\affiliation{Univ. Claude Bernard Lyon1, CNRS, Institut Lumi\`ere Mati\`ere, 69622 Villeurbanne, France}

\author{Alice Nicolas}
\email{alice.nicolas@cea.fr}
\affiliation{Univ. Grenoble Alps, CNRS, LTM, 38000 Grenoble, France}

\date{\today}
\begin{abstract}
To understand cell migration, it is crucial to gain knowledge on how cells exert and integrate forces on/from their environment. A quantity of prime interest for biophysicists interested in cell movements modeling is the intracellular stresses. Up to now, three different methods have been proposed to calculate it, they are all in the regime of the thin plate approximation. Two are based on solving the mechanical equilibrium  equation inside the cell material (Monolayer Stress Microscopy, and Bayesian Inference Stress Microscopy) and one is based on the continuity of displacement at the cell/substrate interface (Intracellular Stress Microscopy). We show here using 3D FEM modeling that these techniques do not calculate the same quantities (as was previously assumed), the first techniques calculate the sum of the active and resistive stresses within the cell, whereas the last one only calculate the resistive component. Combining these techniques should in principle permit to get access to the active stress alone.
\end{abstract}

\pacs{10xxx}
\keywords{biological physics, mechanobiology, traction force microscopy, intracellular stress microscopy, elasticity}

\maketitle

\section{Introduction}

Cell motility is  at the core of both many physiological processes (such as embryogenesis, wound healing...) and pathological processes such as metastasis in cancer \cite{Friedl2009,VanHelvert2018a}. In order to move, cells need to exert forces on its environment \cite{Hakim2017c}. These forces originate either from cellular acto-myosin contractility or from polymerization forces pushing membranes \cite{Schwarz2013a} (these latter forces are transmitted to the substrate on molecular clutches where actin filaments are connected to the substrates \cite{Chan2008a}).
Getting information on these forces is crucial if one wants to really understand individual as well as collective cell migration.
These forces are now routinely accessible using techniques such as Traction Force Microscopy \cite{dembo99,schwarz02}.
They are often used as a simple direct readout, marker free, of cell contractile activity. However, a closer marker of this activity should be given by the
the internal mechanical active stresses generated by cells, as some of the forces exerted on the plane could theoretically be the result of friction (i.e. passive forces, resulting from cell movements)\cite{Notbohm2016}. We explore in this paper the possibility of getting access to this active stresses examining the different techniques which have been developed for
measuring stresses inside the monolayer.

We make explicit the origin of the stress each of these methods calculates, which indeed differs. We validate our approach using Finite Element Modeling of cells submitted to active forces. We show that these different methods allow to quantify the intracellular stress that resists cell active forces and the total intracellular stress accounting for both the active and the resistive stresses. Here we emphasize their complementarity and we make clear the limitations of these calculations. This article is a companion paper of an experimental usage of intracellular stress calculation \cite{DelanoeAyari22PRL}.

\section{MSM, BISM and ISM calculate different intracellular stresses}\label{sec:stresses}
\begin{figure}[!h]
    \centering
    \includegraphics[width=8.5cm]{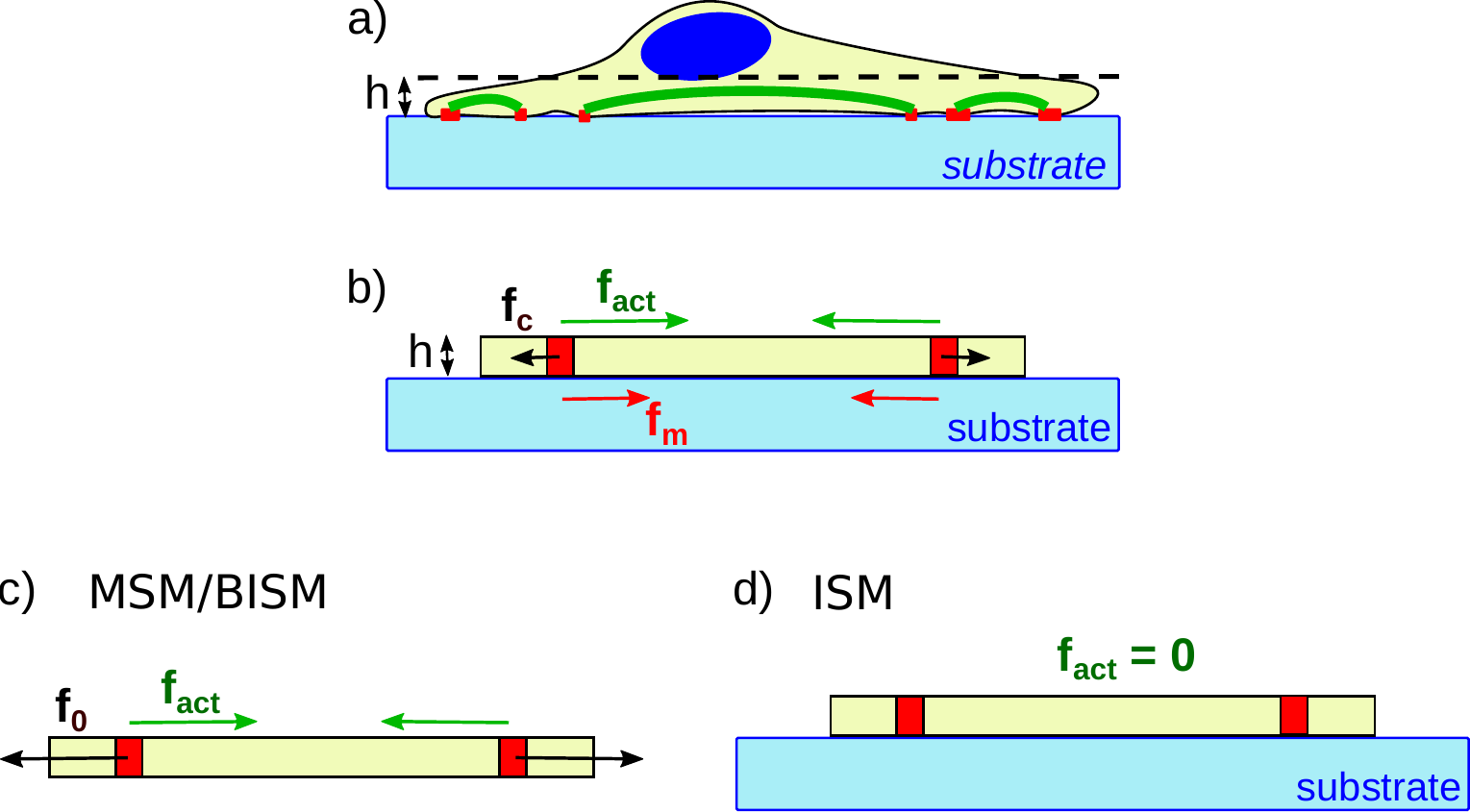}
    \caption{a) Schematics of a contractile cell: the cell body is submitted to internal stresses from,  e.g., acto-myosin filaments (in green) that are transmitted to the substrate through cell adhesions (in red). $h$ is the thickness of the layer where the stresses transmitted to the substrate are generated. b) Mechanical equilibrium at an adhesion point: $\vec{f}_{act}$ is the active force generated in the cell that reaches the adhesion, $\vec{f}_c$ is the  resistance opposed by the cell material, and $\vec{f}_m$ is the force transmitted to the substrate. c-d) Reference states for the calculation of the intracellular stress in (c) MSM and BISM, or (d) in ISM. In c), $\vec{f}_0$ is the  resistance of the cell body in the absence of adhesion.} \label{fig:schema}
    \end{figure}

Active forces in adherent cells generate both resistive forces inside the cells and outside, in the substrate underneath (see Fig. \ref{fig:schema}a--b). Taking advantage of the well-defined mechanical properties of the substrate, and following Ref. \cite{wang02}, several methods have been proposed  that infer intracellular mechanical stresses from the measure of the  resistive force field  in the substrate \cite{tambe11,nier16} or the in-plane deformation field at the surface of the substrate \cite{moussus14}.
As we explain below, this allows quantifying the resistive stress from the cell body \cite{moussus14} or the total intracellular stress associated to the active and resistive intracellular forces \cite{tambe11,nier16}.

The original idea of the mechanical approaches is to model cells as materials subjected to internal volume forces, the active  forces mentioned above (acto-myosin contractility/polymerization). When the cells are adhered to a substrate, the internal forces are transmitted to the substrate and deform it.  Assuming that cell colonies as well as single cells can be modeled as a thin plate, the mechanical equilibrium writes  (Fig. \ref{fig:schema}):
\begin{equation}\label{eq:eqmeca}
\vec{f}_{act}+\vec{f}_c - \vec{f}_m=\vec{0}
\end{equation}
with  $\vec{f}_{act}$  the active cellular forces that cells build up following adhesion, $\vec{f}_c$ and $-\vec{f}_m$ respectively the reaction force of the cell body and the resistance of  the deformable substrate opposed to these active forces, all modeled as surface forces because of the thin plate approximation. $\vec{f}_m$ is precisely the traction stress field measured by traction force microscopy (TFM) \cite{schwarz15}. In the present work, our aim is to characterize $\vec{f}_{act}$ and $\vec{f}_c$.

 Eq. (\ref{eq:eqmeca})  can be reformulated in terms of the stress tensors $S_{act}$ and $S_c$:
 \begin{equation}\label{eq:eqmeca2}
h div S_{act}+h div S_c  - \vec{f}_m=\vec{0}
\end{equation}
where $h$ is the thickness of the plate model. $S_{act}$ is the stress tensor that is derived from the internal cellular force generation following cell adhesion. This stress can be addressed by  a gedanken experiment: let's imagine that the cell could be detached without altering its active stress field, $S_{act}$. Then the cell body would contract till a size determined by the balance with the reactive stress $S_0$ the cell body opposes to its contraction (Fig. \ref{fig:schema}c). Thus $S_{act}=-S_0$.
 Finally, $S_c$ is the stress that results from the strain of the adhered cell material in response to  the internal forces $\vec{f}_{act}$. Both $S_{act}$ and $S_c$ measure stresses in cells following cell adhesion. Prestresses preceding cell adhesion are not accessible here.

The original method, the Monolayer Stress Microscopy (MSM) \cite{wang02,tambe11}, addresses the resolution of Eq. (\ref{eq:eqmeca}) by building a stress tensor $S_{tot}$ that gathers both unknown $S_{act}$ and $S_c$ into a single stress tensor $S_{tot}$:
\begin{align}
S_{tot} = S_{act} + S_c \label{eq:Stot}\\
h div S_{tot} = \vec{f}_m \label{eq:MSM}
\end{align}

Eq. (\ref{eq:MSM}) is underdetermined \cite{timoshenko1951}. An additional relationship between the stress components is added by assuming that the cellular material has a linear elastic rheology \cite{tambe11,tambe13}. In line with MSM, Bayesian Inference Stress Microscopy (BISM) was proposed \cite{nier16}. It also solves the equilibrium Eq. (\ref{eq:MSM}) but accounts for the noise in $\vec{f}_m$ and does not assume a rheological model a priori for the cell material. Underdetermination of Eq. (\ref{eq:MSM}) is resolved by using Bayesian inversion and assuming that $S_{tot}$ has a Gaussian distribution. BISM then introduces a regularization step that allows limiting the contribution of noise in the calculated stress tensor. The rheological properties can be inferred a posteriori, by comparing the temporal derivatives of the elastic strain tensor and the spatial gradient of the velocity field in the cell material \cite{nier16}.

Differently, Intracellular Stress Microscopy (ISM) addresses the quantification of the resistive component of the intracellular stress, $S_c$, that opposes the contraction of the adhered cell \cite{moussus14} (Fig. \ref{fig:schema}d). When the cell is modeled as a thin elastic plate, it is straightly obtained by differentiating the displacement field of the neutral plane of the plate \cite{landau}. This approach can be extended to visco-elastic rheology when the cell material behaves like a Maxwell fluid, a rheological behavior that was for instance reported in flowing epithelial monolayers \cite{tlili20}. When the basal surface of the cell material is uniformly adhered to the substrate, either by integrin-mediated adhesions, non specific adhesions or other types of adhesive machinery such as lectins \cite{moussus14a,DelanoeAyari22PRL}, the displacement field of the neutral plane of the plate is identical to the displacement field on the top of the substrate. The resistive stress then writes:
\begin{gather}
S_c
=\left(
\begin{array}{cc}
\sigma_{xx}&\sigma_{xy}\\
\sigma_{xy}&\sigma_{yy}
\end{array}
\right)\nonumber \\
 \mbox{ with } \left\{
\begin{array}{ccc}
\sigma_{xx}&=&\frac{E_c}{1-\nu_c^2}(\frac{\partial u_x}{\partial x}+\nu_c\frac{\partial  u_y}{\partial y}) \\
\sigma_{yy}&=&\frac{E_c}{1-\nu_c^2}(\frac{\partial u_y}{\partial y}+\nu_c\frac{\partial u_x}{\partial x}) \\
\sigma_{xy}&=&\frac{E_c}{2(1+\nu_c)}\left( \frac{\partial  u_x}{\partial y} + \frac{\partial  u_y}{\partial x}\right)
\end{array}
\right.  \label{eq:ISM}
\end{gather}
$(x,y)$ are the in-plane coordinates, $E_c$ and $\nu_c$ are the Young's modulus and the Poisson's ratio of the cell material of thickness $h$, and $u_{x,y}$ are the in-plane components of the displacement field on top of the substrate. The displacement field is measured as in TFM, by the use of fluorescent markers embedded in the substrate. As a consequence of Eq. (\ref{eq:ISM}), implementation of ISM requires to know the Young's modulus of the cell $E_c$ and its Poisson's ratio $\nu_c$ but is independent of the thickness of the contractile plate, $h$.

MSM or BISM and ISM thus do not address the same intracellular stresses.  MSM or BISM calculates the bidimensional total stress tensor $h S_{tot}=h(S_{act}+S_{c})$ (Eq. (\ref{eq:MSM})) while ISM quantifies the Young's modulus-normalized resistive stress tensor $S_{c}/E_c$ (Eq. (\ref{eq:ISM})).

In the following, by using 3D FEM, we compare the two approaches for calculating intracellular stresses and evaluate their consistency. The first approach was tested by using BISM and not MSM, as experimentally, TFM can only provide $\vec{f}_m$ with a non negligible noise level of more (and often much more) than 10\% \cite{sabass08}, and BISM explicitly handles  the noise level in its formulation. In any case, as both MSM and BISM  are based on the same equation (Eq. \ref{eq:MSM}),  they  should provide similar results as already shown by Nier \emph{et al.} \cite{nier16}.

\section{Using 3D FEM to compare the different calculations}

\begin{figure}[h]
\centering
\includegraphics[width=8 cm]{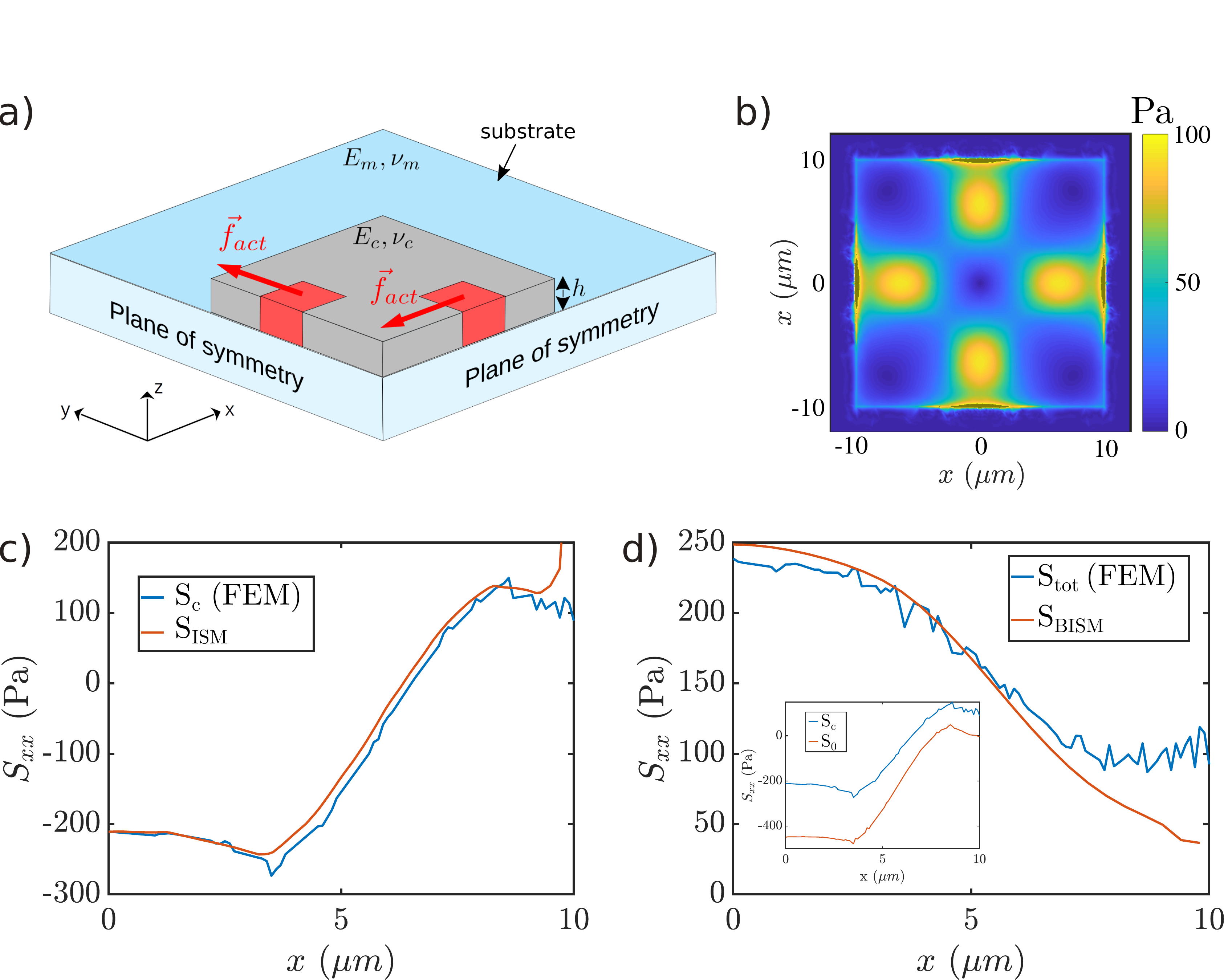}
\caption{FEM calculation of intracellular stresses in an elastic plate ($E_c= 5 kPa$, $\nu_c=0.5$, $h=1 \mu m$) bound to a deformable substrate ($E_m= 1 kPa$, $\nu_m=0.5$). a) Schematics of the numerical experiment. The plate is submitted to contractile and tensile force dipoles respectively along the $x$ and $y$ axis with truncated Gaussian profile (amplitude $1 kPa$, standard deviation $ 2 \mu m$) concentrated in 5 $\mu$m wide squared dots. b) Amplitude of the surface stresses $\vec{f}_m$ on the substrate. c) Comparison of the intracellular resistive stress  $S_c$ calculated with FEM and with ISM ($S_{ISM}$, Eq. (\ref{eq:ISM})). d) Comparison of the total intracellular stress $S_{tot}= S_c -S_0$ and BISM calculation $S_{BISM}$ (Eq. (\ref{eq:MSM}), regularization parameter $L=0.003$). The inset shows the profiles of $S_0 = -S_{act}$ and $S_c$ used in the calculation of $S_{tot}$.} \label{fig:FEM_S}
\end{figure}

Since MSM, BISM and ISM address different intracellular stresses, we built a finite elements simulation in order to calculate these stresses. This approach had already been attempted in Nier \emph{et al.} \cite{nier16}. In this paragraph we analyze this former simulation and show that it can only compute $S_c$ but not $S_{tot}$. We then propose a different simulation that allows obtaining $S_c$ and $S_{tot}$ from 3D FEM.

The simulation in Ref. \cite{nier16} was conceived as follows: the cell monolayer is modeled as a 2D square or disk of either a pure viscous or elastic material. Cellular contractility is modeled as external forces (random dipoles distributed inside the geometry of interest). The substrate is included in the simulation through its interaction with the cells, and enters like a friction term proportional to the velocity (viscous case) or to the displacement  (elastic case, the cells are then firmly attached to the substrate composed of 1D springs). BISM and MSM stresses were calculated by solving $h div S=\vec{t}$, with  $\vec{t}=-\vec{f}_{act}+\xi \vec{u}$ being the forces that act on the cells and $\vec{u}$ the displacement field of the 2D material (note that there is  a minus sign error in the equation used in the supplementary of Ref. \cite{nier16}). Compared to Eq. (\ref{eq:eqmeca}), $\vec{t}$ is therefore the resistive force that opposes the active contraction, $\vec{t}=-\vec{f}_{act} + \vec{f}_m = \vec{f}_c$, and $S=S_c$. Thus the modeling proposed in Ref. \cite{nier16} allows to calculate the reactive stress $S_c$ by two means, either directly with differentiating the displacement field (ISM approach, denoted MSMu in \cite{nier16}) or by solving $h div S_c = \vec{f}_c$. Consistently, BISM, MSM and ISM gave very similar results. Figure S9 in Ref. \cite{nier16} shows that the calculated stresses localize identically for all the methods, either in the elastic or in the viscous cases. Their amplitudes nevertheless differ but this indeed comes from different choices of the rheological parameters in between the tests: for instance, for the viscous case, first and second viscosities are taken equal in the FEM simulation ($\eta=\eta '$) leading to an equivalent Poisson's ratio of 0.25 while the equivalent Poisson's ratio is taken at 0.5 for MSM or ISM; in the same way, the Young's modulus of the cells chosen for ISM and MSM differs from those chosen for  BISM (ISM: $E_c=1kPa$, $\nu_c=0.5$, and $h$ not given, but the only value available is $5 \mu m$ leading to $hE_c=5kPa\cdot \mu m$; for  MSM, $E_c=10kPa$,  $\nu_c= 0.5$, and $h=5 \mu m$, MSM will then compare very well with the FEM simulation in the elastic case as its result only depends on the value of $\nu_c$, and both are taken equal; BISM, in the elastic case, $hE_c=100kPa\cdot \mu m$ or $hE_c=10kPa\cdot \mu m$ and $\nu_c=0.5$). The reason for the failure of the previous simulation to address both $S_c$ and $S_{tot}$ and model a true experiment comes from the fact that $S_{tot}$  is only meaningful when the cells are adhered to a substrate. Otherwise, as detailed in Section \ref{sec:stresses}, $S_{tot}=0$ as the intracellular resistive stress $S_c$ balances the active stress $S_{act}$: in the absence of anchorage to a substrate, $\vec{f}_{act}+\vec{f}_c=\vec{0}$ (Fig. \ref{fig:schema}b).

To solve this issue, we thus proposed a model where the cell (or equivalently the cell colony) is modeled as thin plate uniformly bound to the substrate (Fig. \ref{fig:FEM_S}a). The dimensions of the thin plate were chosen so that the deformation field is fairly uniform in the thickness of the cellular material (square elastic sheet of size 30$\times$30 $\mu m^2$ and $1 \mu m$ in thickness). We focused on the elastic case, with Young's modulus $E_c=5 kPa$ and a Poisson's ratio $\nu_c =0.5$. The thin plate is sitting on top of an elastic gel (the substrate) which is modeled as a thick elastic parallelepiped (size 200$\times$200$\times$100 $\mu m^3$ in (x, y, z), with Young's modulus $E_m=1 kPa$ and a Poisson's ratio $\nu_m =0.5$). A contractile dipole is positioned along the x-axis, composed of Gaussian forces of amplitude $1 kPa$  and width $\sigma$  adjusted between 0.25 and 2 $\mu m$.  A tensile dipole is set on the y-axis with the same amplitude and width.

\section{Details on the robustness of ISM and BISM calculations}

\subsection{Calculation methods}
The 3D FEM calculation provides the displacement field at the interface between the cell and the substrate. This displacement field was used to calculate the intracellular stresses $S_c$ and $S_{tot}$ using ISM and BISM as would be done with experimental data \cite{DelanoeAyari22PRL}. The sampling was chosen following Shannon criterion:  the displacement field was interpolated on a sampled regular grid with a frequency more than twice the maximal frequency obtained from the FEM calculation. The resistive stress field $S_c$ was obtained from ISM, by differentiating the in-plane displacement field retrieved from the 3D FEM calculation using a Sobel approximation of the derivative (Eq. (\ref{eq:ISM})). $S_{tot}$ was obtained from BISM by solving Eq. (\ref{eq:MSM}). Traction stresses $\vec{f}_m$ were first calculated using Fast Fourier Transform, following Butler \textit{et al.} \cite{butler02}. We took $\nu_m=0.49$ for the calculations. The total intracellular stress $S_{tot}$ was then calculated following Ref. \cite{nier16}. As it is quite demanding on computer memory, we used a grid of 50 $\times$ 50 pixels to calculate the stress, which enables a rather fast computation, so as to perform many different tests in a reasonable amount of time. Boundary conditions were enforced in the prior to correspond to the surface forces $\vec{f}_m$ at the edges of the cell. The hyperparameter ensuring $\sigma_{xy}=\sigma_{yx}$, was set to $10^3$ as was done in Ref. \cite{nier16}.

\subsection{Choosing the regularization parameter in BISM}

\begin{figure}[!h]
    \centering
    \includegraphics[width = 8.5cm]{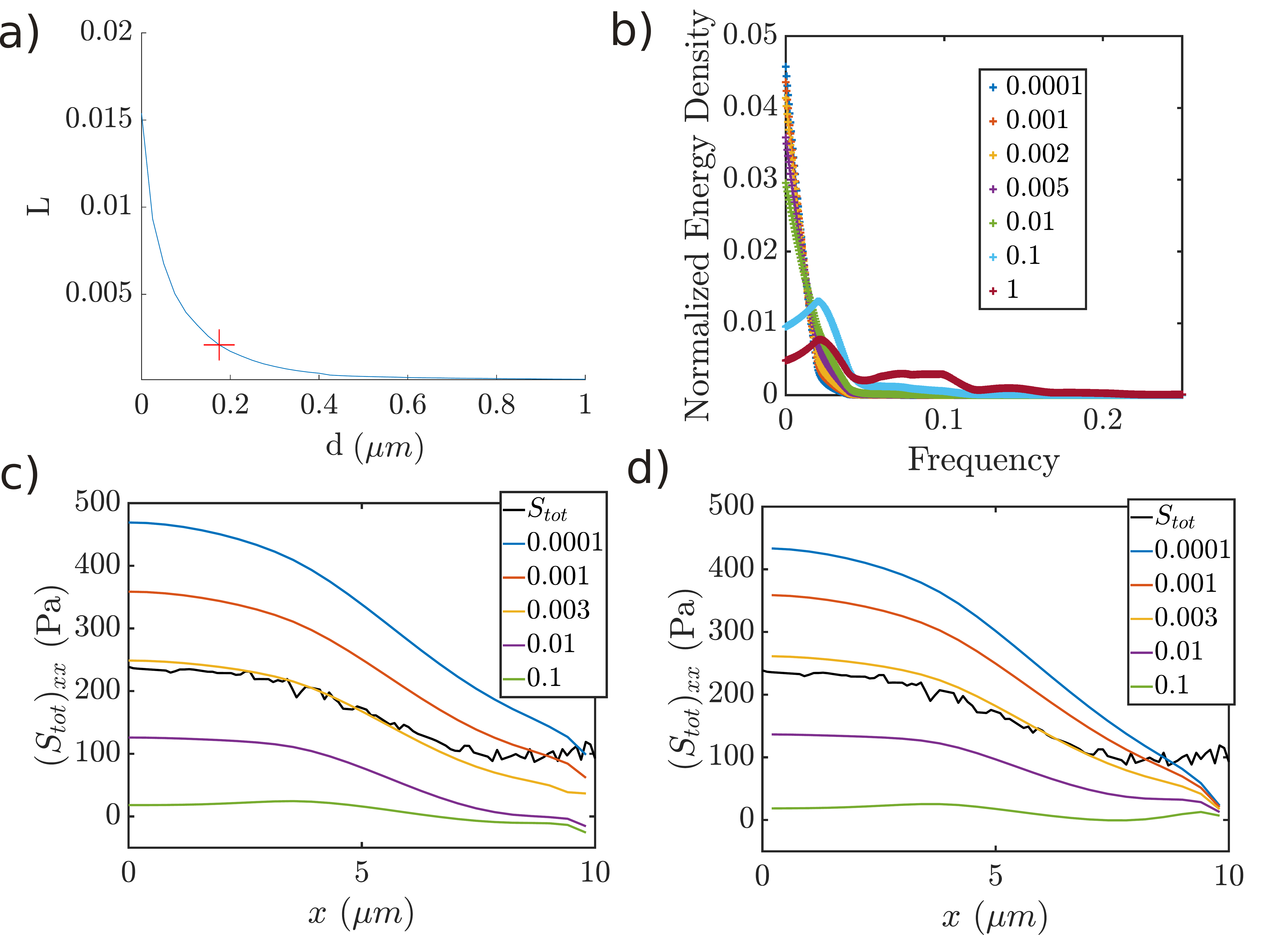}
    \caption{ Understanding BISM.  a) L values calculated with Eq. (\ref{eq:Lvalue}), with s being the noise level outside of the cell quantified in a mask at a distance $d$ from the plate boundaries. The optimum for L is chosen at the maximum of curvature in the curve. b) Increasing the regularization parameter in BISM calculation filters the low frequencies. The colors refer to values of the regulation parameter. c) Boundary conditions are given by the surface forces $\vec{f}_m$ at the edges of the plate. b) Zero stress is assumed at the edges of the plate, as in Ref. \cite{nier16}. The FEM calculation is shown in black. The other colors refer to values of the regularization parameter.}\label{fig:BISM_rob}
    \end{figure}

As detailed in Ref. \cite{caboussat12}, the choice of the optimal parameter for equation form like Eq. (\ref{eq:MSM}) is far from obvious. It is to be noted that the L-curve criterion is not consistent with the Morozov discrepancy principle here (which states that Eq. (\ref{eq:MSM}) can not be solved with a better accuracy than the noise on $\vec{f}_m$), as it gives a dominant weight to the accuracy of the equilibrium equation Eq. (\ref{eq:MSM}), omitting that the right-hand term, $\vec{f}_m$ is a noisy, inaccurate, data. We thus chose  to calculate the regularization parameter $L$ using the $\chi^2$ estimate \cite{schwarz02} which considers the noise level of the right-hand term in Eq. (\ref{eq:MSM}). In BISM, this criterion expresses as $L=\ell^2 s^2/s_0^2$, with $\ell$ the size of the grid sampling for the calculation of $S_{tot}$, $s$ the standard deviation of the noise of $\vec{f}_m$, and $s_0$ the standard deviation of the calculated stress $h S_{tot}$ \cite{nier16}. Since $s_0$ is unknown, an additional criterion is required. Based on Eq. (\ref{eq:MSM}), we estimated $s_0\simeq \ell s_1$, with $s_1$ the standard deviation of $\vec{f}_m$. Then $L$ is simply obtained from $\vec{f}_m$ stress field distribution and the quantification of its noise level out of the cell boundaries:
\begin{equation}\label{eq:Lvalue}
L = s^2/s_1^2
\end{equation}
To calculate the noise, we used the values of the surface forces $\vec{f}_m$ outside of the cell boundaries (it should be zero if the calculation was perfect, which of course is not the case.) However,  the calculation appeared trickier  than for real data coming from experiments. Here, the noise level sharply decreased with distance from the plate. Thus defining the proper position of the boundary appeared mandatory to ensure that the captured noise is not the spread force signal unavoidable with finite element calculation but still is representative of FEM-induced noise. This issue is specific to FEM calculation and is not met in experimental cases where the noise around the cell is fairly uniform (see \cite{DelanoeAyari22PRL}). To this end, noise statistics was quantified out of the cell in regions  whose distance to cell edges was varied. $L$ values were then obtained with Eq. (\ref{eq:Lvalue}) in dependence on this distance (Fig. \ref{fig:BISM_rob}a). $L$ was considered optimal at the maximal curvature of this decreasing curve as it is the place of best compromise between attenuated force signal and maximal noise level. For the values modeled in Fig. \ref{fig:FEM_S}a, we obtained a regularization parameter  $L=0.002$ which consistently corresponds to the best choice for the regularization parameter compared to FEM
calculation of $S_{tot}$ (Fig. \ref{fig:FEM_S}d).

\subsection{Effect of noise in the calculation of $S_{tot}$ and $S_c$}
Noise strongly impacts the calculation of the force field in TFM. This problem was addressed by using a Bayesian approach \cite{dembo96}, a regularization scheme \cite{schwarz02} or a filtering in the Fourier space \cite{butler02}. These regularization schemes were shown to filter high frequencies \cite{ambrosi06, sabass08}. Noise issues keep also critical in the calculation of the intracellular stresses and we questioned how noise impacts ISM and BISM calculations. ISM is based on the derivative of the displacement field. It is therefore very sensitive to high frequency noise. A filtering is applied by the use of the Sobel approximation in the calculation of the gradients.
We showed in a companion paper \cite{DelanoeAyari22PRL} that experimentally, the dispersions of $div S_c$ and $\vec{f}_m$ are similar. The fact that $div S_c$ does not show many points with high amplitude out of the fit line shows that ISM is not altered by high frequency noise compared to TFM. Differently,  BISM calculation is based on  the integration of the surface force field $\vec{f}_m$. A perturbation in $\vec{u}$ with wave vector $\vec{q}$ results in a perturbation of the stress tensor $\Delta S_{tot}$ proportional to $1/q$. Low frequency noise thus strongly alters the value of $S_{tot}$. And indeed, the regularization scheme in BISM calculation damps these low frequencies (Fig. \ref{fig:BISM_rob}b).  Thus in this context, BISM is expected to be very sensitive to the boundary conditions.

\subsection{Effect of boundaries conditions on BISM calculations}
A proper choice of the boundary conditions also appeared to be critical for the success of BISM calculation. $S_{tot}$  was either calculated when assuming zero stress at the edge of the thin plate or when fixing the boundary stress with the surface forces at the edge of the plate: $S_{tot}\cdot \vec{n}= \vec{f}_m$, with $\vec{n}$ the normal to the edge of the plate.  Only the appropriate boundary conditions brought the BISM curve close to the FEM curve (Fig. \ref{fig:BISM_rob}c--d).

\section{BISM and ISM are consistent with FEM}
Stresses calculated with ISM and BISM approaches were compared to the FEM calculation. While $S_c$ is a direct output of the FEM stress tensor calculation,
 $S_{tot}$ was calculated as the difference between  the resistive stress tensors  of the adhered plate ($S_c$) and of the non adhered plate ($ S_0=-S_{act}$) as described in Sec. \ref{sec:stresses}. It was compared to BISM calculation whose value of the regularization parameter was chosen based on the noise level of $\vec{f}_m$.

As shown on Fig. \ref{fig:FEM_S}c, $S_{ISM}$ compared well with $S_c$ in consistence with the thin plate assumption. Similarly, BISM did reconstruct $S_{tot}$ by using appropriate boundary conditions and regularization parameter (Fig. \ref{fig:FEM_S}d), both parameters having an important impact on the stress calculation as was detailed above.

\section{Relationship between div S and f}
\begin{figure}[h]
\centering
\includegraphics[width=8.5 cm]{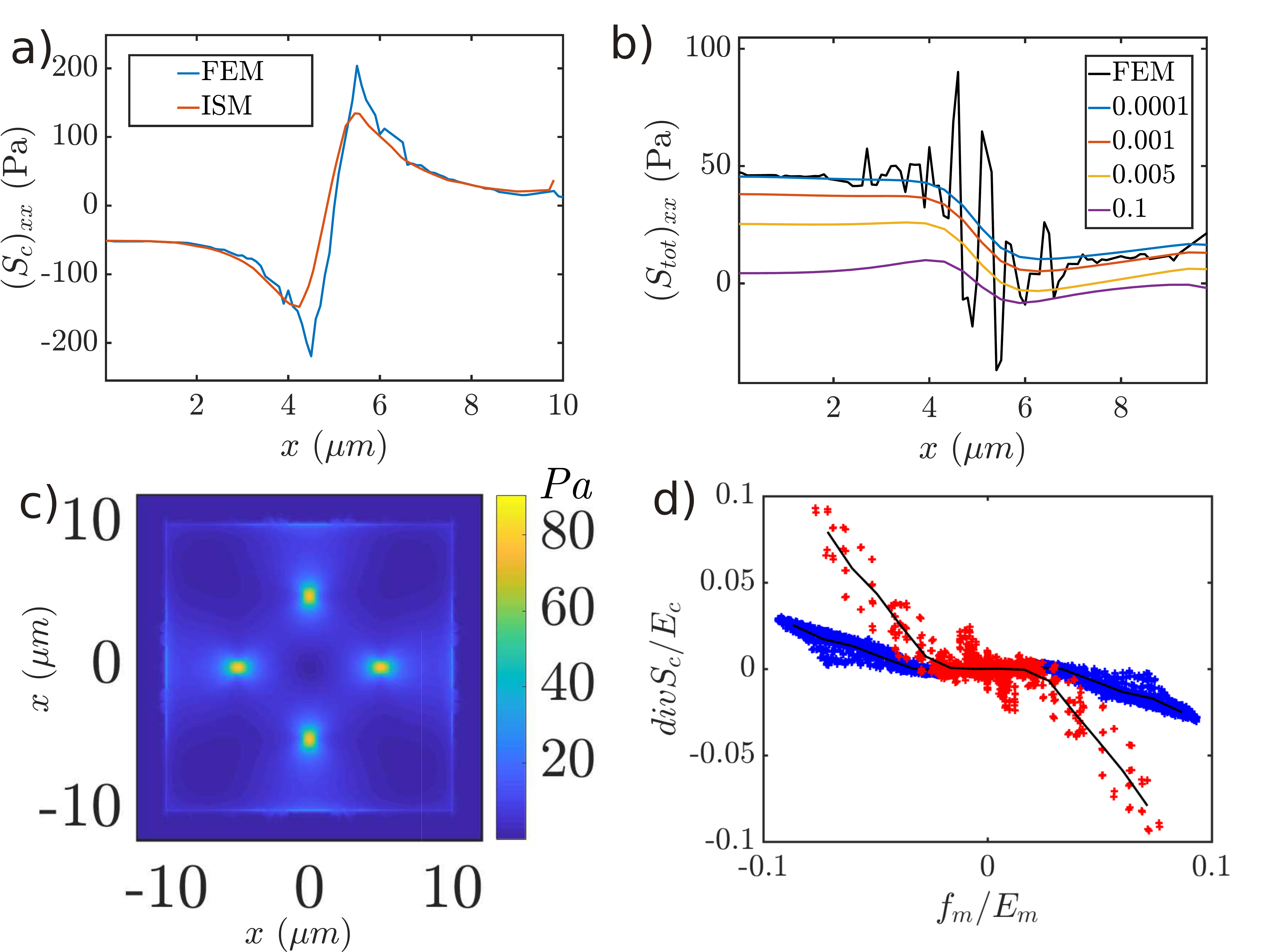}
\caption{Comparison of a) $S_{ISM}$ and b) $S_{BISM}$ with $S_c$ and $S_{tot}$ calculated with FEM simulations for a uniformly adhered plate subjected to a local force field $\vec{f}_{act}$ of Gaussian distribution with similar design as in Fig. \ref{fig:FEM_S}, but of smaller size (force patch of $1\mu m$, $\sigma = 0.25 \mu m$). Again, ISM and BISM well account for the values of $S_c$ (blue line) and $S_{tot}$ (black line). Influence of the value of regularization parameter on the shape of $S_{BISM}$ is shown in (b) (values are listed in the legend).  c) Surface forces $\vec{f}_m$. d) Correlation of $div S_c$ and $\vec{f}_m$.  The sampling size is imposed by the mesh size and is smaller than the width of the Gaussian field (blue: force patch of $5 \mu m$,  $\sigma = 2 \mu m$; red: force patch of $1 \mu m$, $\sigma = 0.25 \mu m$). The black lines are bin averages of the blue or red points.} \label{fig:divS_FEM}
\end{figure}

In \cite{DelanoeAyari22PRL}, we experimentally evidenced a linear relationship between $div(S_c)$ and $\vec{f}_m$, which entails another linear relationship  between $S_c$ and $S_{tot}$. We showed that these linear relationships could only be observed if the sizes of the adhesive active areas were smaller than the resolution of our analysis (\textit{i.e.} 400nm at best).
Here our patches are necessarily above the resolution of our grid. But, we did try to run our modelling on a smaller Gaussian adhesive patch of $1 \mu m$ in size. Results are presented on Fig. \ref{fig:divS_FEM}.
Again BISM and ISM are nicely recovered (Fig. \ref{fig:divS_FEM}a and b) but we  do not recover the linear relationship between $div(S_c)$ and $\vec{f}_m$ (Fig. \ref{fig:divS_FEM}d). This is normal
as  the size of the patches are necessarily larger than the sampling size fixed by the mesh. However, when reducing these sizes, the relationship tends toward more linear (compare the blue and red lines).

\section{Case of localized adhesion }

\begin{figure}[h]
\includegraphics[width=8.5cm]{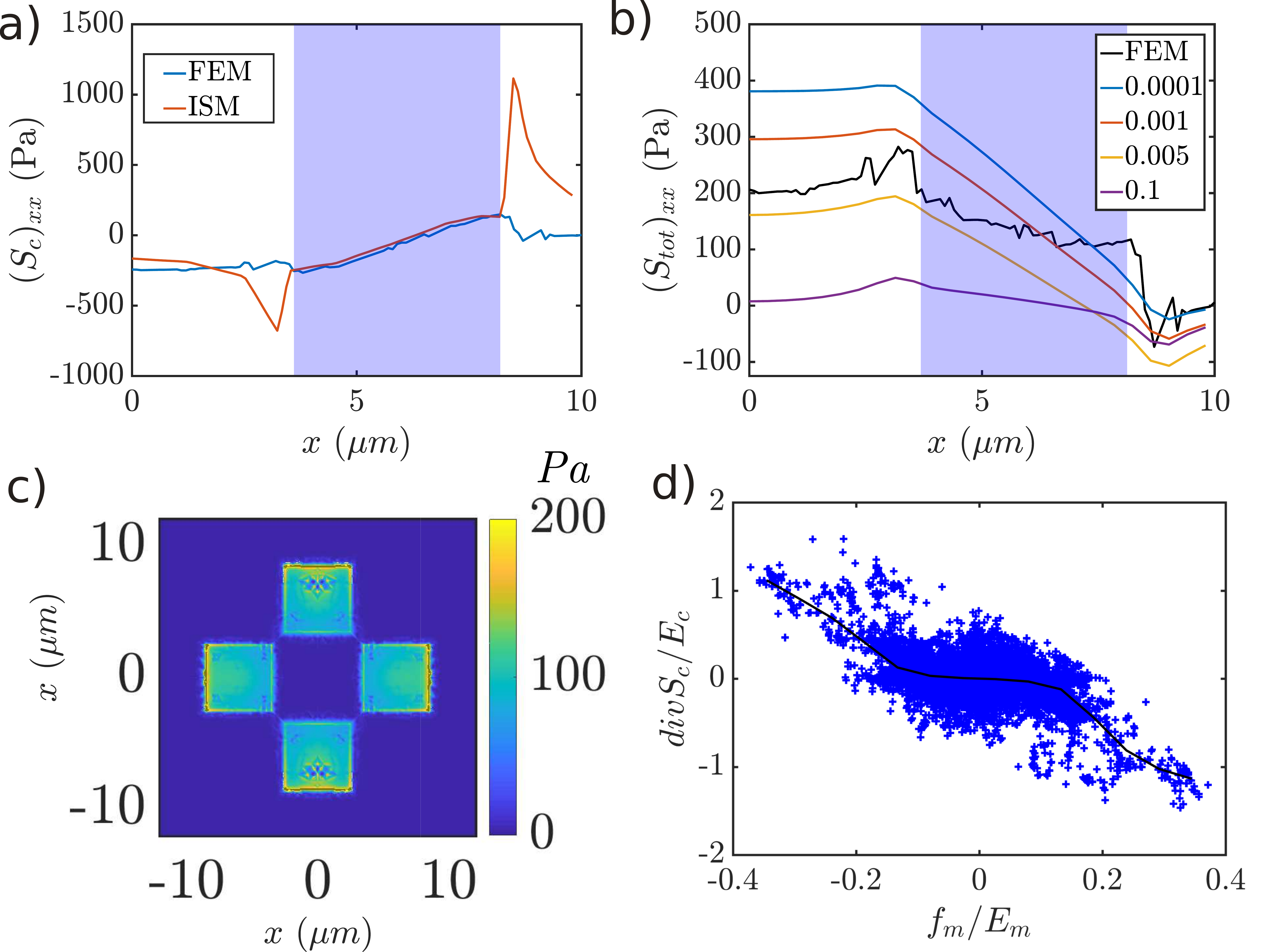}
\caption{Comparison of a) $S_{ISM}$ and b) $S_{BISM}$ with $S_c$ and $S_{tot}$ calculated with FEM simulations for a plate only adhered in the patches where active surface forces apply. The characteristics of the active force field is identical to Fig. 2 in the main text (force patch of $5 \mu m$, $\sigma = 2 \mu m$). Here ISM is no more relevant out of the adhered patch and BISM also fails to represent $S_{tot}$ (shown with a black line). The adhesive area is delimited by the light purple area in both a and b. Influence of the value of the regularization parameter on the shape of $S_{BISM}$ is shown in (b) (values are listed in the legend). c) Amplitude map of the surface forces $\vec{f}_m$ for a plate that is only adhered through the local patches where the active forces $\vec{f}_{act}$ are generated (force patch of $5 \mu m$, $\sigma = 2 \mu m$). c) Correlation between $divS_c$ and $\vec{f}_m$ for the locally adhered plate (same parameters as in (a)).}\label{fig:noadh}
\end{figure}

We then examined the case where the cells do not adhere everywhere.  We tested how this situation would impact the correlation between $div S_{ISM}$ and $\vec{f}_m$. It should be noted that out of the areas where the cell is adhered,  $S_{ISM}$ differs from $S_c$ as its calculation is based on the deformation of the substrate which now differs from the deformation of the cell. Using 3D FEM simulations, we observed that the surface forces $\vec{f}_m$ concentrate in the regions of adhesion only (Fig. \ref{fig:noadh}a). Would intracellular stress generation occur out of the adhered areas, the surface forces $\vec{f}_m$ would change in amplitude but of course they would still concentrate into the adhesive patches. As a result, it could not be inferred by the calculation of the total stress $hS_{tot}$ would not be able to show it as Eq. (\ref{eq:MSM}) does not contain any information on the location of the stress generators. Thus, as expected, the calculation of $S_{tot}$ is less robust when the cell is not continuously adhered (Fig. \ref{fig:noadh}). $div S_{ISM}$ and $\vec{f}_m$ showed a correlation that resembles the one obtained for adherent cells in the presence of force generators of width larger than the sampling size, although it appeared more noisy (compare Fig. \ref{fig:divS_FEM}d  and Fig. \ref{fig:noadh}d). Thus the experimental observation of a linear relationship between $div S_{ISM}$ and $\vec{f}_m$  may not be sufficient to conclude on the adhesive interaction of the cells with the substrate, the measure of transmitted surface forces $\vec{f}_m$ that emerge from the noise being more conclusive. The intracellular stresses generated in regions where the level of force transmission to the substrate is low thus remain difficult to characterize by these mechanical approaches.

\section{Conclusion}
We proposed here a set of 3D FEM simulations to test and validate intracellular stress calculations that are done by  two independent techniques, ISM and BISM, which addresses two different stresses.
We showed that BISM enables to measure the total stress inside the cells, while ISM retrieves the resistive component of the stress. We delineated the framework within which these techniques provide consistent information:  for both techniques, the calculation infers relevant stresses only at the locations where the cells are adherent; concerning BISM, a proper quantification of the noise level to select the optimal regularization parameter and a proper definition of the boundary conditions are mandatory. Within this well-defined framework, we have shown that both approaches bring valuable and complementary information on intracellular stresses which then allow the retrieval of the active part of the intracellular stress. Taking advantage of  this knowledge, it is now possible to analyze intracellular stresses in real experiments. This is what we have done in a companion paper \cite{DelanoeAyari22PRL}.

\begin{acknowledgments}
The authors are indebted to P. Marcq for the provision of the BISM calculation code. This work was initiated by very fruitful discussions with E. Mazza, L. Filotto, P. Silberzan and T. Vourc'h. H. D. and A. N. are grateful to them.  A. N. acknowledges the support by ANR-12-JSVE05-0008.
\end{acknowledgments}

\bibliographystyle{apsrev4-2}
%



\begin{thebibliography}{24}%
\makeatletter
\providecommand \@ifxundefined [1]{%
 \@ifx{#1\undefined}
}%
\providecommand \@ifnum [1]{%
 \ifnum #1\expandafter \@firstoftwo
 \else \expandafter \@secondoftwo
 \fi
}%
\providecommand \@ifx [1]{%
 \ifx #1\expandafter \@firstoftwo
 \else \expandafter \@secondoftwo
 \fi
}%
\providecommand \natexlab [1]{#1}%
\providecommand \enquote  [1]{``#1''}%
\providecommand \bibnamefont  [1]{#1}%
\providecommand \bibfnamefont [1]{#1}%
\providecommand \citenamefont [1]{#1}%
\providecommand \href@noop [0]{\@secondoftwo}%
\providecommand \href [0]{\begingroup \@sanitize@url \@href}%
\providecommand \@href[1]{\@@startlink{#1}\@@href}%
\providecommand \@@href[1]{\endgroup#1\@@endlink}%
\providecommand \@sanitize@url [0]{\catcode `\\12\catcode `\$12\catcode
  `\&12\catcode `\#12\catcode `\^12\catcode `\_12\catcode `\%12\relax}%
\providecommand \@@startlink[1]{}%
\providecommand \@@endlink[0]{}%
\providecommand \url  [0]{\begingroup\@sanitize@url \@url }%
\providecommand \@url [1]{\endgroup\@href {#1}{\urlprefix }}%
\providecommand \urlprefix  [0]{URL }%
\providecommand \Eprint [0]{\href }%
\providecommand \doibase [0]{https://doi.org/}%
\providecommand \selectlanguage [0]{\@gobble}%
\providecommand \bibinfo  [0]{\@secondoftwo}%
\providecommand \bibfield  [0]{\@secondoftwo}%
\providecommand \translation [1]{[#1]}%
\providecommand \BibitemOpen [0]{}%
\providecommand \bibitemStop [0]{}%
\providecommand \bibitemNoStop [0]{.\EOS\space}%
\providecommand \EOS [0]{\spacefactor3000\relax}%
\providecommand \BibitemShut  [1]{\csname bibitem#1\endcsname}%
\let\auto@bib@innerbib\@empty
\bibitem [{\citenamefont {Friedl}\ and\ \citenamefont
  {Gilmour}(2009)}]{Friedl2009}%
  \BibitemOpen
  \bibfield  {author} {\bibinfo {author} {\bibfnamefont {P.}~\bibnamefont
  {Friedl}}\ and\ \bibinfo {author} {\bibfnamefont {D.}~\bibnamefont
  {Gilmour}},\ }\href {https://doi.org/10.1038/nrm2720} {\bibfield  {journal}
  {\bibinfo  {journal} {Nat. Rev. Mol. Cell Biol.}\ }\textbf {\bibinfo {volume}
  {10}},\ \bibinfo {pages} {445} (\bibinfo {year} {2009})}\BibitemShut
  {NoStop}%
\bibitem [{\citenamefont {{Van Helvert}}\ \emph {et~al.}(2018)\citenamefont
  {{Van Helvert}}, \citenamefont {Storm},\ and\ \citenamefont
  {Friedl}}]{VanHelvert2018a}%
  \BibitemOpen
  \bibfield  {author} {\bibinfo {author} {\bibfnamefont {S.}~\bibnamefont {{Van
  Helvert}}}, \bibinfo {author} {\bibfnamefont {C.}~\bibnamefont {Storm}},\
  and\ \bibinfo {author} {\bibfnamefont {P.}~\bibnamefont {Friedl}},\ }\href
  {https://doi.org/10.1038/s41556-017-0012-0} {\bibfield  {journal} {\bibinfo
  {journal} {Nat. Cell Biol.}\ }\textbf {\bibinfo {volume} {20}},\ \bibinfo
  {pages} {8} (\bibinfo {year} {2018})}\BibitemShut {NoStop}%
\bibitem [{\citenamefont {Hakim}\ and\ \citenamefont
  {Silberzan}(2017)}]{Hakim2017c}%
  \BibitemOpen
  \bibfield  {author} {\bibinfo {author} {\bibfnamefont {V.}~\bibnamefont
  {Hakim}}\ and\ \bibinfo {author} {\bibfnamefont {P.}~\bibnamefont
  {Silberzan}},\ }\href {https://doi.org/10.1088/1361-6633/aa65ef} {\bibfield
  {journal} {\bibinfo  {journal} {Rep. Prog. Phys.}\ }\textbf {\bibinfo
  {volume} {80}},\ \bibinfo {pages} {076601} (\bibinfo {year}
  {2017})}\BibitemShut {NoStop}%
\bibitem [{\citenamefont {Schwarz}\ and\ \citenamefont
  {Safran}(2013)}]{Schwarz2013a}%
  \BibitemOpen
  \bibfield  {author} {\bibinfo {author} {\bibfnamefont {U.~S.}\ \bibnamefont
  {Schwarz}}\ and\ \bibinfo {author} {\bibfnamefont {S.~A.}\ \bibnamefont
  {Safran}},\ }\href {https://doi.org/10.1103/RevModPhys.85.1327} {\bibfield
  {journal} {\bibinfo  {journal} {Rev. Mod. Phys.}\ }\textbf {\bibinfo {volume}
  {85}},\ \bibinfo {pages} {1327} (\bibinfo {year} {2013})}\BibitemShut
  {NoStop}%
\bibitem [{\citenamefont {Chan}\ and\ \citenamefont {Odde}(2008)}]{Chan2008a}%
  \BibitemOpen
  \bibfield  {author} {\bibinfo {author} {\bibfnamefont {C.~E.}\ \bibnamefont
  {Chan}}\ and\ \bibinfo {author} {\bibfnamefont {D.~J.}\ \bibnamefont
  {Odde}},\ }\href {https://doi.org/10.1126/science.1163595} {\bibfield
  {journal} {\bibinfo  {journal} {Science}\ }\textbf {\bibinfo {volume}
  {322}},\ \bibinfo {pages} {1687} (\bibinfo {year} {2008})}\BibitemShut
  {NoStop}%
\bibitem [{\citenamefont {Dembo}\ \emph {et~al.}(1999)\citenamefont {Dembo},
  \citenamefont {Wang},\ and\ \citenamefont {Y}}]{dembo99}%
  \BibitemOpen
  \bibfield  {author} {\bibinfo {author} {\bibfnamefont {M.}~\bibnamefont
  {Dembo}}, \bibinfo {author} {\bibfnamefont {Y.~L.}\ \bibnamefont {Wang}},\
  and\ \bibinfo {author} {\bibnamefont {Y}},\ }\href
  {https://doi.org/10.1016/S0006-3495(99)77386-8} {\bibfield  {journal}
  {\bibinfo  {journal} {Biophys. J.}\ }\textbf {\bibinfo {volume} {76}},\
  \bibinfo {pages} {2307} (\bibinfo {year} {1999})}\BibitemShut {NoStop}%
\bibitem [{\citenamefont {Schwarz}\ \emph {et~al.}(2002)\citenamefont
  {Schwarz}, \citenamefont {Balaban}, \citenamefont {Riveline}, \citenamefont
  {Bershadsky}, \citenamefont {Geiger},\ and\ \citenamefont
  {Safran}}]{schwarz02}%
  \BibitemOpen
  \bibfield  {author} {\bibinfo {author} {\bibfnamefont {U.~S.}\ \bibnamefont
  {Schwarz}}, \bibinfo {author} {\bibfnamefont {N.}~\bibnamefont {Balaban}},
  \bibinfo {author} {\bibfnamefont {D.}~\bibnamefont {Riveline}}, \bibinfo
  {author} {\bibfnamefont {A.}~\bibnamefont {Bershadsky}}, \bibinfo {author}
  {\bibfnamefont {B.}~\bibnamefont {Geiger}},\ and\ \bibinfo {author}
  {\bibfnamefont {S.}~\bibnamefont {Safran}},\ }\href
  {http://linkinghub.elsevier.com/retrieve/pii/S000634950273909X} {\bibfield
  {journal} {\bibinfo  {journal} {Biophys. J.}\ }\textbf {\bibinfo {volume}
  {83}},\ \bibinfo {pages} {1380} (\bibinfo {year} {2002})}\BibitemShut
  {NoStop}%
\bibitem [{\citenamefont {Notbohm}\ \emph {et~al.}(2016)\citenamefont
  {Notbohm}, \citenamefont {Banerjee}, \citenamefont {Utuje}, \citenamefont
  {Gweon}, \citenamefont {Jang}, \citenamefont {Park}, \citenamefont {Shin},
  \citenamefont {Butler}, \citenamefont {Fredberg},\ and\ \citenamefont
  {Marchetti}}]{Notbohm2016}%
  \BibitemOpen
  \bibfield  {author} {\bibinfo {author} {\bibfnamefont {J.}~\bibnamefont
  {Notbohm}}, \bibinfo {author} {\bibfnamefont {S.}~\bibnamefont {Banerjee}},
  \bibinfo {author} {\bibfnamefont {K.~J.}\ \bibnamefont {Utuje}}, \bibinfo
  {author} {\bibfnamefont {B.}~\bibnamefont {Gweon}}, \bibinfo {author}
  {\bibfnamefont {H.}~\bibnamefont {Jang}}, \bibinfo {author} {\bibfnamefont
  {Y.}~\bibnamefont {Park}}, \bibinfo {author} {\bibfnamefont {J.}~\bibnamefont
  {Shin}}, \bibinfo {author} {\bibfnamefont {J.~P.}\ \bibnamefont {Butler}},
  \bibinfo {author} {\bibfnamefont {J.~J.}\ \bibnamefont {Fredberg}},\ and\
  \bibinfo {author} {\bibfnamefont {M.~C.}\ \bibnamefont {Marchetti}},\ }\href
  {https://doi.org/10.1016/j.bpj.2016.05.019} {\bibfield  {journal} {\bibinfo
  {journal} {Biophys. J.}\ }\textbf {\bibinfo {volume} {110}},\ \bibinfo
  {pages} {2729} (\bibinfo {year} {2016})}\BibitemShut {NoStop}%
\bibitem [{\citenamefont {Delano{\"{e}}-Ayari}\ \emph
  {et~al.}(2022)\citenamefont {Delano{\"{e}}-Ayari}, \citenamefont
  {Bouchonville}, \citenamefont {Cour{\c{c}}on},\ and\ \citenamefont
  {Nicolas}}]{DelanoeAyari22PRL}%
  \BibitemOpen
  \bibfield  {author} {\bibinfo {author} {\bibfnamefont {H.}~\bibnamefont
  {Delano{\"{e}}-Ayari}}, \bibinfo {author} {\bibfnamefont {N.}~\bibnamefont
  {Bouchonville}}, \bibinfo {author} {\bibfnamefont {M.}~\bibnamefont
  {Cour{\c{c}}on}},\ and\ \bibinfo {author} {\bibfnamefont {A.}~\bibnamefont
  {Nicolas}},\ }\href@noop {} {\bibfield  {journal} {\bibinfo  {journal}
  {submitted to Phys. Rev. Lett.}\ } (\bibinfo {year} {2022})}\BibitemShut
  {NoStop}%
\bibitem [{\citenamefont {Wang}\ \emph {et~al.}(2002)\citenamefont {Wang},
  \citenamefont {Toli-N{\o}rrelykke}, \citenamefont {Chen}, \citenamefont
  {Mijailovich}, \citenamefont {Butler}, \citenamefont {Fredberg},\ and\
  \citenamefont {Stamenovi\'c}}]{wang02}%
  \BibitemOpen
  \bibfield  {author} {\bibinfo {author} {\bibfnamefont {N.}~\bibnamefont
  {Wang}}, \bibinfo {author} {\bibfnamefont {I.~M.}\ \bibnamefont
  {Toli-N{\o}rrelykke}}, \bibinfo {author} {\bibfnamefont {J.}~\bibnamefont
  {Chen}}, \bibinfo {author} {\bibfnamefont {S.~M.}\ \bibnamefont
  {Mijailovich}}, \bibinfo {author} {\bibfnamefont {J.~P.}\ \bibnamefont
  {Butler}}, \bibinfo {author} {\bibfnamefont {J.~J.}\ \bibnamefont
  {Fredberg}},\ and\ \bibinfo {author} {\bibfnamefont {D.}~\bibnamefont
  {Stamenovi\'c}},\ }\href
  {https://doi.org/10.1152/AJPCELL.00269.2001/ASSET/IMAGES/LARGE/H00320869109.JPEG}
  {\bibfield  {journal} {\bibinfo  {journal} {Am. J. Physiol. Cell Physiol.}\
  }\textbf {\bibinfo {volume} {282}},\ \bibinfo {pages} {606} (\bibinfo {year}
  {2002})}\BibitemShut {NoStop}%
\bibitem [{\citenamefont {Tambe}\ \emph {et~al.}(2011)\citenamefont {Tambe},
  \citenamefont {{Corey Hardin}}, \citenamefont {Angelini}, \citenamefont
  {Rajendran}, \citenamefont {Park}, \citenamefont {Serra-Picamal},
  \citenamefont {Zhou}, \citenamefont {Zaman}, \citenamefont {Butler},
  \citenamefont {Weitz}, \citenamefont {Fredberg},\ and\ \citenamefont
  {Trepat}}]{tambe11}%
  \BibitemOpen
  \bibfield  {author} {\bibinfo {author} {\bibfnamefont {D.~T.}\ \bibnamefont
  {Tambe}}, \bibinfo {author} {\bibfnamefont {C.}~\bibnamefont {{Corey
  Hardin}}}, \bibinfo {author} {\bibfnamefont {T.~E.}\ \bibnamefont
  {Angelini}}, \bibinfo {author} {\bibfnamefont {K.}~\bibnamefont {Rajendran}},
  \bibinfo {author} {\bibfnamefont {C.~Y.}\ \bibnamefont {Park}}, \bibinfo
  {author} {\bibfnamefont {X.}~\bibnamefont {Serra-Picamal}}, \bibinfo {author}
  {\bibfnamefont {E.~H.}\ \bibnamefont {Zhou}}, \bibinfo {author}
  {\bibfnamefont {M.~H.}\ \bibnamefont {Zaman}}, \bibinfo {author}
  {\bibfnamefont {J.~P.}\ \bibnamefont {Butler}}, \bibinfo {author}
  {\bibfnamefont {D.~A.}\ \bibnamefont {Weitz}}, \bibinfo {author}
  {\bibfnamefont {J.~J.}\ \bibnamefont {Fredberg}},\ and\ \bibinfo {author}
  {\bibfnamefont {X.}~\bibnamefont {Trepat}},\ }\href
  {https://doi.org/10.1038/nmat3025} {\bibfield  {journal} {\bibinfo  {journal}
  {Nat. Mater.}\ }\textbf {\bibinfo {volume} {10}},\ \bibinfo {pages} {469}
  (\bibinfo {year} {2011})}\BibitemShut {NoStop}%
\bibitem [{\citenamefont {Nier}\ \emph {et~al.}(2016)\citenamefont {Nier},
  \citenamefont {Jain}, \citenamefont {Lim}, \citenamefont {Ishihara},
  \citenamefont {Ladoux},\ and\ \citenamefont {Marcq}}]{nier16}%
  \BibitemOpen
  \bibfield  {author} {\bibinfo {author} {\bibfnamefont {V.}~\bibnamefont
  {Nier}}, \bibinfo {author} {\bibfnamefont {S.}~\bibnamefont {Jain}}, \bibinfo
  {author} {\bibfnamefont {C.~T.}\ \bibnamefont {Lim}}, \bibinfo {author}
  {\bibfnamefont {S.}~\bibnamefont {Ishihara}}, \bibinfo {author}
  {\bibfnamefont {B.}~\bibnamefont {Ladoux}},\ and\ \bibinfo {author}
  {\bibfnamefont {P.}~\bibnamefont {Marcq}},\ }\href
  {https://doi.org/10.1016/j.bpj.2016.03.002} {\bibfield  {journal} {\bibinfo
  {journal} {Biophys. J.}\ }\textbf {\bibinfo {volume} {110}},\ \bibinfo
  {pages} {1625} (\bibinfo {year} {2016})}\BibitemShut {NoStop}%
\bibitem [{\citenamefont {Moussus}\ \emph
  {et~al.}(2014{\natexlab{a}})\citenamefont {Moussus}, \citenamefont {{der
  Loughian}}, \citenamefont {Fuard}, \citenamefont {Cour{\c{c}}on},
  \citenamefont {Gulino-Debrac}, \citenamefont {Delano{\"{e}}-Ayari},\ and\
  \citenamefont {Nicolas}}]{moussus14}%
  \BibitemOpen
  \bibfield  {author} {\bibinfo {author} {\bibfnamefont {M.}~\bibnamefont
  {Moussus}}, \bibinfo {author} {\bibfnamefont {C.}~\bibnamefont {{der
  Loughian}}}, \bibinfo {author} {\bibfnamefont {D.}~\bibnamefont {Fuard}},
  \bibinfo {author} {\bibfnamefont {M.}~\bibnamefont {Cour{\c{c}}on}}, \bibinfo
  {author} {\bibfnamefont {D.}~\bibnamefont {Gulino-Debrac}}, \bibinfo {author}
  {\bibfnamefont {H.}~\bibnamefont {Delano{\"{e}}-Ayari}},\ and\ \bibinfo
  {author} {\bibfnamefont {A.}~\bibnamefont {Nicolas}},\ }\href
  {https://doi.org/10.1039/c3sm52318g} {\bibfield  {journal} {\bibinfo
  {journal} {Soft Matter}\ }\textbf {\bibinfo {volume} {10}},\ \bibinfo {pages}
  {2414} (\bibinfo {year} {2014}{\natexlab{a}})}\BibitemShut {NoStop}%
\bibitem [{\citenamefont {Schwarz}\ and\ \citenamefont
  {Soin{\'{e}}}(2015)}]{schwarz15}%
  \BibitemOpen
  \bibfield  {author} {\bibinfo {author} {\bibfnamefont {U.~S.}\ \bibnamefont
  {Schwarz}}\ and\ \bibinfo {author} {\bibfnamefont {J.~R.}\ \bibnamefont
  {Soin{\'{e}}}},\ }\href {https://doi.org/10.1016/j.bbamcr.2015.05.028}
  {\bibfield  {journal} {\bibinfo  {journal} {Biochim. Biophys. Acta, Mol. Cell
  Res.}\ }\textbf {\bibinfo {volume} {1853}},\ \bibinfo {pages} {3095}
  (\bibinfo {year} {2015})}\BibitemShut {NoStop}%
\bibitem [{\citenamefont {Timoshenko}\ and\ \citenamefont
  {Goodier}(1951)}]{timoshenko1951}%
  \BibitemOpen
  \bibfield  {author} {\bibinfo {author} {\bibfnamefont {S.}~\bibnamefont
  {Timoshenko}}\ and\ \bibinfo {author} {\bibfnamefont {J.~N.}\ \bibnamefont
  {Goodier}},\ }\href@noop {} {\bibfield  {journal} {\bibinfo  {journal} {Inc.
  New York}\ } (\bibinfo {year} {1951})}\BibitemShut {NoStop}%
\bibitem [{\citenamefont {Tambe}\ \emph {et~al.}(2013)\citenamefont {Tambe},
  \citenamefont {Croutelle}, \citenamefont {Trepat}, \citenamefont {Park},
  \citenamefont {Kim}, \citenamefont {Millet}, \citenamefont {Butler},\ and\
  \citenamefont {Fredberg}}]{tambe13}%
  \BibitemOpen
  \bibfield  {author} {\bibinfo {author} {\bibfnamefont {D.~T.}\ \bibnamefont
  {Tambe}}, \bibinfo {author} {\bibfnamefont {U.}~\bibnamefont {Croutelle}},
  \bibinfo {author} {\bibfnamefont {X.}~\bibnamefont {Trepat}}, \bibinfo
  {author} {\bibfnamefont {C.~Y.}\ \bibnamefont {Park}}, \bibinfo {author}
  {\bibfnamefont {J.~H.}\ \bibnamefont {Kim}}, \bibinfo {author} {\bibfnamefont
  {E.}~\bibnamefont {Millet}}, \bibinfo {author} {\bibfnamefont {J.~P.}\
  \bibnamefont {Butler}},\ and\ \bibinfo {author} {\bibfnamefont {J.~J.}\
  \bibnamefont {Fredberg}},\ }\href
  {https://doi.org/10.1371/journal.pone.0055172} {\bibfield  {journal}
  {\bibinfo  {journal} {PLoS One}\ }\textbf {\bibinfo {volume} {8}},\ \bibinfo
  {pages} {e55172} (\bibinfo {year} {2013})}\BibitemShut {NoStop}%
\bibitem [{\citenamefont {Landau}\ and\ \citenamefont
  {Lifshitz}(1970)}]{landau}%
  \BibitemOpen
  \bibfield  {author} {\bibinfo {author} {\bibfnamefont {L.~D.}\ \bibnamefont
  {Landau}}\ and\ \bibinfo {author} {\bibfnamefont {E.~M.}\ \bibnamefont
  {Lifshitz}},\ }\href@noop {} {\emph {\bibinfo {title} {{Theory of
  elasticity}}}},\ \bibinfo {edition} {2nd}\ ed.,\ \bibinfo {series} {Course of
  Theoretical Physics}, Vol.~\bibinfo {volume} {7}\ (\bibinfo  {publisher}
  {Pergamon Press, Oxford},\ \bibinfo {year} {1970})\BibitemShut {NoStop}%
\bibitem [{\citenamefont {Tlili}\ \emph {et~al.}(2020)\citenamefont {Tlili},
  \citenamefont {Durande}, \citenamefont {Gay}, \citenamefont {Ladoux},
  \citenamefont {Graner},\ and\ \citenamefont {Delano{\"{e}}-Ayari}}]{tlili20}%
  \BibitemOpen
  \bibfield  {author} {\bibinfo {author} {\bibfnamefont {S.}~\bibnamefont
  {Tlili}}, \bibinfo {author} {\bibfnamefont {M.}~\bibnamefont {Durande}},
  \bibinfo {author} {\bibfnamefont {C.}~\bibnamefont {Gay}}, \bibinfo {author}
  {\bibfnamefont {B.}~\bibnamefont {Ladoux}}, \bibinfo {author} {\bibfnamefont
  {F.}~\bibnamefont {Graner}},\ and\ \bibinfo {author} {\bibfnamefont
  {H.}~\bibnamefont {Delano{\"{e}}-Ayari}},\ }\href
  {https://doi.org/10.1103/physrevlett.125.088102} {\bibfield  {journal}
  {\bibinfo  {journal} {Phys. Rev. Lett.}\ }\textbf {\bibinfo {volume} {125}},\
  \bibinfo {pages} {88102} (\bibinfo {year} {2020})}\BibitemShut {NoStop}%
\bibitem [{\citenamefont {Moussus}\ \emph
  {et~al.}(2014{\natexlab{b}})\citenamefont {Moussus}, \citenamefont {{der
  Loughian}}, \citenamefont {Fuard}, \citenamefont {Cour\c{c}on}, \citenamefont
  {Gulino-Debrac}, \citenamefont {Delano\"e-Ayari},\ and\ \citenamefont
  {Nicolas}}]{moussus14a}%
  \BibitemOpen
  \bibfield  {author} {\bibinfo {author} {\bibfnamefont {M.}~\bibnamefont
  {Moussus}}, \bibinfo {author} {\bibfnamefont {C.}~\bibnamefont {{der
  Loughian}}}, \bibinfo {author} {\bibfnamefont {D.}~\bibnamefont {Fuard}},
  \bibinfo {author} {\bibfnamefont {M.}~\bibnamefont {Cour\c{c}on}}, \bibinfo
  {author} {\bibfnamefont {D.}~\bibnamefont {Gulino-Debrac}}, \bibinfo {author}
  {\bibfnamefont {H.}~\bibnamefont {Delano\"e-Ayari}},\ and\ \bibinfo {author}
  {\bibfnamefont {A.}~\bibnamefont {Nicolas}},\ }\href
  {https://doi.org/10.1039/c4sm01066c} {\bibfield  {journal} {\bibinfo
  {journal} {Soft Matter}\ }\textbf {\bibinfo {volume} {10}},\ \bibinfo {pages}
  {7683} (\bibinfo {year} {2014}{\natexlab{b}})}\BibitemShut {NoStop}%
\bibitem [{\citenamefont {Sabass}\ \emph {et~al.}(2008)\citenamefont {Sabass},
  \citenamefont {Gardel}, \citenamefont {Waterman-Storer},\ and\ \citenamefont
  {Schwarz}}]{sabass08}%
  \BibitemOpen
  \bibfield  {author} {\bibinfo {author} {\bibfnamefont {B.}~\bibnamefont
  {Sabass}}, \bibinfo {author} {\bibfnamefont {M.~L.}\ \bibnamefont {Gardel}},
  \bibinfo {author} {\bibfnamefont {C.~M.}\ \bibnamefont {Waterman-Storer}},\
  and\ \bibinfo {author} {\bibfnamefont {U.~S.}\ \bibnamefont {Schwarz}},\
  }\href {https://doi.org/10.1529/biophysj.107.113670} {\bibfield  {journal}
  {\bibinfo  {journal} {Biophys. J.}\ }\textbf {\bibinfo {volume} {94}},\
  \bibinfo {pages} {207} (\bibinfo {year} {2008})}\BibitemShut {NoStop}%
\bibitem [{\citenamefont {Butler}\ \emph {et~al.}(2002)\citenamefont {Butler},
  \citenamefont {Toli{\'{c}}-N{\o}rrelykke}, \citenamefont {Fabry},\ and\
  \citenamefont {Fredberg}}]{butler02}%
  \BibitemOpen
  \bibfield  {author} {\bibinfo {author} {\bibfnamefont {J.~P.}\ \bibnamefont
  {Butler}}, \bibinfo {author} {\bibfnamefont {I.}~\bibnamefont
  {Toli{\'{c}}-N{\o}rrelykke}}, \bibinfo {author} {\bibfnamefont
  {B.}~\bibnamefont {Fabry}},\ and\ \bibinfo {author} {\bibfnamefont {J.~J.}\
  \bibnamefont {Fredberg}},\ }\href
  {http://view.ncbi.nlm.nih.gov/pubmed/11832345
  http://ajpcell.physiology.org/content/282/3/C595.short} {\bibfield  {journal}
  {\bibinfo  {journal} {Am. J. Physiol. Cell Physiol.}\ }\textbf {\bibinfo
  {volume} {282}},\ \bibinfo {pages} {C595} (\bibinfo {year}
  {2002})}\BibitemShut {NoStop}%
\bibitem [{\citenamefont {Caboussat}\ and\ \citenamefont
  {Glowinski}(2012)}]{caboussat12}%
  \BibitemOpen
  \bibfield  {author} {\bibinfo {author} {\bibfnamefont {A.}~\bibnamefont
  {Caboussat}}\ and\ \bibinfo {author} {\bibfnamefont {R.}~\bibnamefont
  {Glowinski}},\ }\href@noop {} {\bibfield  {journal} {\bibinfo  {journal} {J.
  Comput. Math.}\ }\textbf {\bibinfo {volume} {30}},\ \bibinfo {pages} {354}
  (\bibinfo {year} {2012})}\BibitemShut {NoStop}%
\bibitem [{\citenamefont {Dembo}\ \emph {et~al.}(1996)\citenamefont {Dembo},
  \citenamefont {Oliver}, \citenamefont {Ishihara},\ and\ \citenamefont
  {Jacobson}}]{dembo96}%
  \BibitemOpen
  \bibfield  {author} {\bibinfo {author} {\bibfnamefont {M.}~\bibnamefont
  {Dembo}}, \bibinfo {author} {\bibfnamefont {T.}~\bibnamefont {Oliver}},
  \bibinfo {author} {\bibfnamefont {A.}~\bibnamefont {Ishihara}},\ and\
  \bibinfo {author} {\bibfnamefont {K.}~\bibnamefont {Jacobson}},\ }\href
  {https://doi.org/10.1016/S0006-3495(96)79767-9} {\bibfield  {journal}
  {\bibinfo  {journal} {Biophys. J.}\ }\textbf {\bibinfo {volume} {70}},\
  \bibinfo {pages} {2008} (\bibinfo {year} {1996})}\BibitemShut {NoStop}%
\bibitem [{\citenamefont {Ambrosi}(2006)}]{ambrosi06}%
  \BibitemOpen
  \bibfield  {author} {\bibinfo {author} {\bibfnamefont {D.}~\bibnamefont
  {Ambrosi}},\ }\href
  {http://citeseerx.ist.psu.edu/viewdoc/download?doi=10.1.1.104.4980{\&}rep=rep1{\&}type=pdf}
  {\bibfield  {journal} {\bibinfo  {journal} {SIAM J. Appl. Math.}\ }\textbf
  {\bibinfo {volume} {66}},\ \bibinfo {pages} {2049} (\bibinfo {year}
  {2006})}\BibitemShut {NoStop}%
\end{thebibliography}

\end{document}